\input harvmac.tex

\def\np#1#2#3{Nucl. Phys. {\bf B#1} (#2) #3}
\def\pl#1#2#3{Phys. Lett. {\bf #1B} (#2) #3}
\def\prl#1#2#3{Phys. Rev. Lett. {\bf #1} (#2) #3}
\def\prd#1#2#3{Phys. Rev. {\bf D#1} (#2) #3}

\def\jhep#1#2#3{JHEP {\bf#1}(#2) #3}
\def\jmp#1#2#3{J. Math Phys. {\bf #1} (#2) #3}

%%%%%%%%%%%%%%%%%%%%%  Rublenye bukvy   %%%%%%%%%%%%%%%%%%%%%%%%
\def\IB{\relax\hbox{$\inbar\kern-.3em{\rm B}$}}
\def\IC{\relax\hbox{$\inbar\kern-.3em{\rm C}$}}
\def\ID{\relax\hbox{$\inbar\kern-.3em{\rm D}$}}
\def\IE{\relax\hbox{$\inbar\kern-.3em{\rm E}$}}
\def\IF{\relax\hbox{$\inbar\kern-.3em{\rm F}$}}
\def\IG{\relax\hbox{$\inbar\kern-.3em{\rm G}$}}
\def\IGa{\relax\hbox{${\rm I}\kern-.18em\Gamma$}}
\def\IH{\relax{\rm I\kern-.18em H}}
\def\IK{\relax{\rm I\kern-.18em K}}
\def\IL{\relax{\rm I\kern-.18em L}}
\def\IP{\relax{\rm I\kern-.18em P}}
\def\IR{\relax{\rm I\kern-.18em R}}
\def\IZ{\relax\ifmmode\mathchoice
{\hbox{\cmss Z\kern-.4em Z}}{\hbox{\cmss Z\kern-.4em Z}}
{\lower.9pt\hbox{\cmsss Z\kern-.4em Z}}
{\lower1.2pt\hbox{\cmsss Z\kern-.4em Z}}\else{\cmss Z\kern-.4em Z}\fi}

\def\II{\relax{\rm I\kern-.18em I}}

%%%%%%%%%%%%%%%%%%%%% Calligraphic letters  %%%%%%%%%%%%%%%%%%%%%

\def\CN {{\cal N}}

\def\CR {{\cal R}}

\def\CW {{\cal W}}

\def\Br {{\bf r}}
%%%%%%%%%%%%%%%%%%%%%%%%%% Derivatives  %%%%%%%%%%%%%%%%%%%%%%%%

%%%%%%%%%%%%%%%%%%%% letters with bar %%%%%%%%%%%%%%%%%%%%%%%%%%

\def\zb {\bar{z}}

%%%%%%%%%%%%%%%%%%%%%%%%%%% Math symbols %%%%%%%%%%%%%%%%%%%%%%%

\def\Tr{{\rm Tr}}

\def\Hom{{\rm Hom}}
\def\dim{{\rm dim}}

\def\liet{{\underline{\bf t}}}

\def\inbar{\,\vrule height1.5ex width.4pt depth0pt}
\font\cmss=cmss10 \font\cmsss=cmss10 at 7pt

%%%%%%%%%%%%%%%%%%%%%%%%%%%%%%%
%REFERENCES

%AdS duals of field theories
\lref\juanads{J. Maldacena, ``The Large-N Limit of Superconformal
Field Theories and Supergravity'', hep-th/9712200}
\lref\ksorb{S. Kachru and E. Silverstein, ``4d Conformal
Field Theories and Strings on Orbifolds'', hep-th/9802183}

%D-branes on orbifolds
\lref\dmquiver{M.R. Douglas and G. Moore, ``D-branes, Quivers and
ALE Instantons'', hep-th/9603167}
\lref\dgmorb{M.R. Douglas, B.R. Greene and D.R. Morrison,
``Orbifold Resolution by D-branes'', \np{505}{1997}{84}; hep-th/9704151}
\lref\dougegs{M.R. Douglas, ``Enhanced Gauge Symmetry
in M(atrix) theory'', \jhep{007}{1997}{004}; hep-th/9612126}
\lref\jmorb{C.V. Johnson and R.C. Myers, ``Aspects of
Type IIB Theory on ALE Spaces'', \prd{55}{1997}{6382}, hep-th/9610140}
\lref\gipol{E.G. Gimon and J. Polchinski, ``Consistency Conditions for
Orientifolds and D Manifolds'', \prd{54}{1996}{1667}, hep-th/9601038}
\lref\polten{J. Polchinski, ``Tensors from K3 Orientifolds'',
\prd{55}{1997}{6423}; hep-th/9606165}
\lref\frbranes{D.-E. Diaconescu, M.R. Douglas and Jaume Gomis,
``Fractional Branes and Wrapped Branes'', hep-th/9712230}

%Geometric engineering
\lref\kmvgeom{S. Katz, P. Mayr and C. Vafa, ``Mirror Symmetry and Exact
Solution of 4D $\CN=2$ Gauge Theories -- I'', hep-th/9706110}
\lref\branegeom{H. Ooguri and C. Vafa, ``Geometry of $\CN=1$ dualities
in four dimensions'', \np{500}{1997}{62}; hep-th/9702180}

\lref\iba{L.E. Ib\'a\~nez,``A Chiral D=4, N=1 String Vacuum with
a Finite Low Energy Effective Field Theory,'' hep-th/9802103}

\lref\kleb{S. Gubser and I. Klebanov,``Absorption by Branes
and Schwinger Terms in the World Volume Theory,'' Phys. Lett. {\bf
B413} (1997)
41.}

%brane construction of GT
\lref\mfour{E. Witten, ``Solutions of Four-dimensional Field Theories
via M-theory'', \np{500}{1997}{3}; hep-th/9703166}

%enhanced gauge symmetry in type II
\lref\aspegs{P.S. Aspinwall, ``Enhanced Gauge Symmetries and K3
Surfaces'',
\pl{357}{1995}{329}; hep-th/9507012}
\lref\bsvegs{M. Bershadsky, V. Sadov and C. Vafa, ``D-strings on
D-manifolds'',
\np{463}{1996}{398}}
%Finite theories
\lref\hps{S.~Hamidi, J.Patera, J.~Schwarz, ``Chiral Two-Loop Finite
Supersymmetric Theories'', Phys. Lett. {\bf 141B} (1984) 349}
\lref\twoloop{D.~R.~T.~Jones and L.~Mezincescu,
``The Beta Function in Supersymmetric
Yang-Mills Theory'', Phys. Lett. {\bf B137} (1984) 242\semi
P.~West, ``The Yukawa Beta-Function in $\CN =1$ Rigid Supersymmetric
Theories'', \pl{137}{1984}{371}\semi
A.~Parkes and P.~West, ``Finiteness  in Rigid Supersymmetric Theories''
\pl{138}{1984}{99}\semi
D.~R.~T.~Jones and L.~Mezincescu, ``The Chiral Anomaly and a Class of
Two-Loop Finite Supersymmetric Gauge Theories'', \pl{138}{1984}{293}}
%susy
\lref\shifvai{M.~Shifman and A.~Vainshtein, ``Solution of the Anomaly
Puzzle
in Susy Gauge Theories and the Wilson Operator Expansion'',
\np{277}{1986}{456} \semi
``On Holomorphic Dependence and Infrared Effects in Supersymmetric
Gauge Theories'', \np{359}{1991}{571}}
\lref\robmat{R.~G.~Leigh, M.J.~Strassler,
``Exactly Marginal Operators and Duality in Four Dimensional N=1
Supersymmetric Gauge Theory'', Nucl. Phys. {\bf B}447 (1995) 95}

%Discrete gauge symmetries in 4d
\lref\kwdiscrete{L.M. Krauss and F. Wilczek, ``Discrete Gauge Symmetries
in Continuum Theories'', \prl{62}{1989}{1221}}

%N=4 field theories
\lref\sixteen{N. Seiberg, ``Notes on Theories with 16 Supercharges'',
hep-th/9705117}

%McKay correspondence
\lref\reidrev{M. Reid, ``McKay correspondence'', alg-geom/9702016}

%Group theory
\lref\suthree{W.M. Fairbanks, T. Fulton and W.H. Klink, ``Finite and
Disconnected Subgroups of $SU(3)$ and their Application to the
Elementary Particle Spectrum'', \jmp{5}{1964}{1038}}
\lref\sosix{W.~Plesken and M.~Pohst, Math. Comp. {\bf 31} (1977) 552}
%%%%%%%%%

\Title{ \vbox{\baselineskip12pt\hbox{hep-th/9803015}
\hbox{HUTP-98/A015}
\hbox{ITEP-TH-15/98}}}
{\vbox{
 \centerline{On Conformal Field Theories in Four Dimensions}}}
\medskip
\centerline{Albion  Lawrence $^2$, Nikita Nekrasov $^{1,2}$ and
Cumrun Vafa
 $^2$}

\vskip 0.5cm
\centerline{$^{1}$ Institute of Theoretical and Experimental
Physics,
117259, Moscow, Russia}
\centerline{$^2$ Lyman Laboratory of Physics,
Harvard University, Cambridge, MA 02138}

\medskip
\bigskip
\noindent
Extending recent work of Kachru and Silverstein, we consider
``orbifolds''
of 4-dimensional $\CN=4$ $SU(n)$ super-Yang-Mills theories with
respect to
discrete subgroups of the $SU(4)$ $R$-symmetry which act
nontrivially on the
gauge group.  We show that for every discrete
subgroup of $SU(4)$ there is a canonical choice of imbedding of the
discrete group in the gauge
group which leads to
theories with a vanishing one-loop beta-function.
We conjecture that these give rise to (generically non-supersymmetric)
conformal theories.  The gauge group is $\otimes_i SU(Nn_i)$ where $n_i$
denote the dimension of the irreducible representations of the
corresponding
discrete group; there is also bifundamental
matter, dictated by associated
quiver diagrams.  The interactions can also be read off from
the quiver diagram.  For $SU(3)$ and $SU(2)$ subgroups this
leads to superconformal theories with $\CN=1$ and $\CN=2$ respectively.
In the $\CN=1$ case we prove the vanishing of the
beta functions to two loops.
\medskip
\Date{March 1998}
%\draftmode

\newsec{Introduction}
It has been a longstanding problem in quantum field theories
to obtain non-trivial four dimensional theories with vanishing
beta functions which lead to conformal theories \hps.
With recent progress in
understanding
of supersymmetric gauge theories we have learned how superconformal
theories
can arise in certain cases.  However, not much
progress has been made for the non-supersymmetric case, due in particular
to an absence of non-renormalization theorems.
Very recently, motivated by connections between the gauge systems
on branes and their supergravity realizations at large $N$ \kleb\juanads,
Kachru and Silverstein \ksorb\ (following the discussions in refs
\dmquiver\jmorb\dougegs\dgmorb) considered an ``orbifold''
 construction of gauge theories
in four dimensions (with or without supersymmetry) and argued they
should be related to conformal fixed points (at least in large $N$)\foot{
The $\CN=1$ example considered in \ksorb\ had previously been
considered in connection with finding a finite $\CN=1$ theory in \iba.}.
The aim of this paper is to extend their construction.  We find
a large set of proposed gauge theories in four dimensions, which we
conjecture
will lead to conformal fixed points.  As a first check we show that they
all have vanishing one-loop beta functions for the gauge couplings.

We construct for every discrete subgroup $\Gamma\subset SU(4)$ a
gauge theory
consisting of gauge group $G=\otimes SU(Nn_i)$ where $N$ is an
arbitrary integer and $n_i$ denote the dimensions of the irreducible
representations of $\Gamma$.  The matter and interactions can be read
off from an associated ``quiver'' diagram.  It
consists of one node for each irreducible representation
of $\Gamma$ and fermionic and bosonic
arrows connecting the nodes according to how
the $4$ and $6$ dimensional represenations
of $\Gamma$ (inherited from $SU(4)$) act on each irreducible
representation.  To each bosonic/fermionic arrow from node $i$ to node
$j$ we associate a bifundamental  $(Nn_i,\overline{Nn_j})$
scalar/Weyl fermion  (for $i=j$ this corresponds to an adjoint
representation).
There is a Yukawa coupling for each triangle
on the quiver, consisting of two fermionic arrows and a bosonic arrow;
and quartic scalar interactions for each square on the quiver,
consisting of four bosonic arrows.  The resulting
gauge theory is chiral if and only if $\Gamma$ is a complex
subgroup of $SU(4)$, i.e. if $\bf 4$ and $\bf {\bar 4}$
of $SU(4)$ are inequivalent representations of $\Gamma$.
If $\Gamma\subset SU(3)$ the
resulting theory is an $\CN=1$ theory and if $\Gamma\subset SU(2)$
it is an $\CN=2$ theory. In such
cases we can
use the supersymmery structure to discuss
a reduced quiver (with only one kind of arrow, consisting
of multiplication with $\bf 3$ of $SU(3)$ and $\bf 2$ of $SU(2)$).

If $\Gamma \subset SU(2)$ we obtain an $\CN=2$ theory.  In fact
it is known that this
gives us {\it all} possible superconformal theories
with gauge group $\otimes_i SU(Nn_i)$ and bifundamental
matter \kmvgeom, i.e. those theories which arise from quiver diagrams
corresponding to affine ADE Dynkin diagrams.    Similarly,
we conjecture that for
the $\CN=1,0$ cases the theories we
construct, we
also exhaust the possibilities for superconformal theories with gauge
group $\otimes_i SU(Nn_i)$
and bifundamental matter.    In all cases, we can provide
some geometric intuition for the meaning of the choices
we have made for ``orbifolding'' based on
the geometry of branes imbedded in a string background.

\newsec{Projections preserving conformal invariance}

Kachru and Silverstein suggested  that if one studies $N$ D3-branes
on various orbifolds of $\IR^6$ and follows  the conjecture
of \juanads\ (i.e. for values of $N$, $g_s$, and $M_s$ for which
that picture makes sense), then one finds that one is studying an
orbifold
of $AdS_5 \times S^5$ where the orbifold group acts only on the $S^5$
factor.  Since the conformal group of the field theory is identified
with the isometry group of $AdS_5$, these orbifolds should lead
to conformal field theories on the D3-branes.

\subsec{Projection of the field theory: general story}

For the purposes of understanding the field content, one can simply work
with the perturbative open string theory on the D-branes, as was
done in \dmquiver\jmorb\ for the $\CN=2$ case and in \dgmorb\ for
the $\CN=1$ case.  Since we would like to keep our discussion as
close to field theory as possible, we will abstract the discussion
in \dmquiver\ below and discuss general projections on the fields
of the original $\CN=4$ theory.  Within this abstract
setup, we can ask which projections lead to vanishing
one-loop beta-functions for the gauge coupling.  Of course,
in general the beta functions for other couplings must be checked
as well.

We will start with an $U(n)$ $\CN=4$ field theory.  As is well known
(see for example \sixteen\ for a discussion with references), this
theory has an $Spin(6) = SU(4)$ $R$-symmetry group
(the transverse rotation group of the D3-brane in the perturbative
open string picture, or the rotation group of the $S^5$ in \juanads.)
It contains the gauge bosons $A_{IJ}$ ($I,J=1, \ldots, n$)
which are singlets of $SO(6)$;
adjoint Weyl fermions $\Psi^{\alpha}_{IJ}$ with $\alpha$
in a ${\bf 4}$ of $SU(4)$;
and adjoint scalars $\Phi^{m}_{IJ}$ with $m$ in
the ${\bf 6}$ of $SO(6)$.  We now wish to pick a discrete subgroup
$\Gamma$ of $Spin(6)$ which acts nontrivially on the gauge group;
such an action corresponds to the action of the orbifold group on the
Chan-Paton
factors of the open strings \gipol\dmquiver.  Let $\{\Br_i\}$
be the set of unitary irreducible representations of $\Gamma$.
We can specify the action
by breaking up the indices $I$ into various representations
$\CR_{i} = \IC^{N_{i}} \Br_i$ of $\Gamma$
such that $\sum_{i} N_{i} \dim(\Br_{i}) = n$, where $N_i$
denotes the number of times the representation $\Br_i$ appears
in this decomposition.  Here $\Gamma$ acts
trivially on $\IC^{N_{i}}$.
Note that for projections which break only part of the supersymmetry,
$\Gamma$ will lie in $SU(2)$, preserving $\CN=2$ or $SU(3)$,
preserving $\CN=1$.

We now consider a modified theory whose fields correspond to keeping
$\Gamma$-invariant fields of the original theory, together with the
terms in the action containing them.  This is {\it not} orbifolding
in the
usual sense
of that word as it is not gauging a discrete
symmetry.   However, we sometimes will continue to refer to this
as orbifolding or projecting the original theory; the
justification of this terminology comes from the fact that in string
context such theories arise upon orbifolding string backgrounds.
One may write $U(n)$
adjoint fields which are singlets under the broken R-symmetry
as homomorphisms from $\IC^{n}$ to itself, or as $\IC^{n}\otimes
(\IC^{n})^{\ast}$.
The effect of this projection is easily derived:
$$
\eqalign{
	&\Hom \left( \IC^n, \IC^n\right)^{\Gamma} =
	\bigoplus_{i,j} \Hom \left( \CR_{i},  \CR_{j}
\right)^{\Gamma} = \cr
	&
\bigoplus_{i,j} \left( \IC^{N_{i}}\otimes(\IC^{N_{j}})^{\ast}\otimes
		\Br_i \otimes \Br^{\ast}_j \right)^{\Gamma}
	= \bigoplus_{i} \IC^{N_i} \otimes (\IC^{N_i})^{\ast}\ ,}
$$
where the superscript $\Gamma$ means keeping only the trivial
representations
in the decomposition with respect to the irreps of $\Gamma$.
Thus $R$-charge singlets break up into adjoints of $U(N_i)$;
in particular the unbroken gauge group is
$$
	G_{{\rm proj}} = \otimes_i U(N_i)\ .
$$
The $U(1)$ factors decouple at low energies, so we will in fact consider
$$G_{{\rm proj}} = \otimes_i SU(N_i)\ .$$
The projection for fields carrying $R$-charge is a bit more complicated.
Let us examine fields transforming under the R-symmetry.  For each
(not necessarily irreducible) representation $\CR$ of $\Gamma$
define the coefficients
$a^{\CR}_{ij}$ by the equations
\eqn\defmult{\CR \otimes \Br_{i} = \oplus_{j} a^{\CR}_{ij}
		\Br_{j} }
Then we may describe the projection of an adjoint field with
$R$-charge as
$$
\eqalign{&\left( \CR\otimes \Hom(\IC^{n}, \IC^{n})\right)^{\Gamma}
		= \bigoplus_{i,j} \left( \CR \otimes \CR_i
\otimes \CR^{\ast}_j \right)^{\Gamma} = \cr
		& \bigoplus_{i,j,k} a^{\CR}_{ij} \left(
\CR_j \otimes \CR^{\ast}_{k}\right)^{\Gamma}
		= \bigoplus_{i,j} a^{\CR}_{i,j}
\IC^{N_i} \otimes (\IC^{N_{i}})^{\ast}\ .}
$$
Now let ${\bf 4}, {\bf 6}$ denote representations of $\Gamma$ coming
from the fundamental and antisymmetric
 representations of $SU(4)$.  Since the Weyl fermions transform
according to ${\bf 4}$ and scalars according to ${\bf 6}$ of the
$R$-symmetry group, we have
$a_{ij}^{\bf 4}$ fermions $\Psi^{ij}_{f_{ij}}$ where $f_{ij}$
runs from $1$ to $a_{ij}^{\bf 4}$ and $a_{ij}^{\bf 6}$ scalars
$\Phi^{ij}_{f_{ij}}$ where $f_{ij}$
runs from $1$ to $a_{ij}^{\bf 6}$; all of these lie in the
$(N_i, \bar{N}_j)$ representation of the group (in the scalar case
we can use the reverse arrow contribution from $j$ to $i$ and thus
think of these as complex fields).

Even though one can easily work out the general case,
because we are motivated by the search for conformal theories
we will be interested in a specific case where we choose
the regular representation of $\Gamma$.  In other words we start
with an $U(n)$ gauge group, assume $n =N \vert \Gamma \vert$,
and think of the fundamental representation of $U(n)$ as
decomposing to the space ${\bf C}^N\otimes \{g \} $ with
$g\in \Gamma$.  We consider the action of $\Gamma$
to be on the second index as right multiplication.  As is well
known, this represention decomposes to a direct sum of all
irreducible representions of $\Gamma$ with degeneracy factor
$n_i={\rm dim}{\bf r}_i$ for the ${\bf r}_i$ representation.
In this case the gauge group we obtain is
$$G=\otimes_i SU(Nn_i)$$
where the index $i$ runs over the irreducible representations
of $\Gamma$.  It is a straightforward exercise to show
that the coupling constant $\tau_i$ of the i-th group
(including the theta angle in the usual way) is given by
$$\tau_i={n_i\tau\over |\Gamma|}$$
where $\tau$ is the $\CN=4$ coupling (or
alternatively the type IIB coupling).
Note in particular that
\eqn\coupling{\sum_in_i\tau_i=\tau}

The matter content is naturally summarized by
an associated ``quiver'' diagram.  It
consists of one node for each irreducible representation
of $\Gamma$, and fermionic and bosonic
arrows connecting the nodes according to how
the $4$- and $6$-dimensional representations
of $\Gamma $ (inherited from $SU(4)$) act on each irreducible
representation.  To each bosonic/fermionic arrow from node $i$ to node
$j$ we associate a bifundamental  $(Nn_i,\overline{Nn_j})$
scalar/Weyl fermion  (for $i=j$ this corresponds to an adjoint
representation).
There are Yukawa couplings for each triangle
on the quiver consisting of two fermionic arrows and a bosonic arrow,
and quartic scalar interactions for each square on the quiver
consisting of four bosonic arrows.  The coefficients of each
interaction can be read off by projecting the original $\CN=4$
lagrangian in terms of the fields we have kept.  The Yukawa
couplings are given by
\eqn\yuk{Y = \sum_{ij k} \gamma_{ijk}^{f_{ij}, f_{jk}, f_{ki}} {\Tr}
\Psi_{f_{ij}}^{ij} \Phi_{f_{jk}}^{jk} \Psi_{f_{ki}}^{ki} }
and the quartic $\phi^{4}$ terms are given by
\eqn\qua{V = \sum_{ijkl} \eta^{ijkl}_{f_{ij}, f_{jk}, f_{kl},
f_{li}} {\Tr}
\Phi_{f_{ij}}^{ij}\Phi_{f_{jk}}^{jk}\Phi_{f_{kl}}^{kl}\Phi_{f_{li}}^{li}
\ ,}
where
\eqn\gma{\gamma_{ij k}^{f_{ij}, f_{jk}, f_{ki}} =
\Gamma_{\alpha\beta, m} \left( Y_{f_{ij}}
\right)^{\alpha}_{v_{i}{\bar v}_{j}}
\left( Y_{f_{jk}} \right)^{m}_{v_{j}{\bar v}_{k}} \left( Y_{f_{ki}}
\right)^{\beta}_{v_{k}{\bar v}_{i}}}
and
\eqn\eeta{\eta^{ijkl}_{f_{ij}, f_{jk}, f_{kl}, f_{li}} =
\left(Y_{f_{ij}}\right)^{[ m}_{v_{i}{\bar v}_{j}}
\left(Y_{f_{jk}}\right)^{n]}_{v_{j}{\bar v}_{k}}
\left(Y_{f_{kl}}\right)^{[m}_{v_{k}{\bar v}_{l}}
\left(Y_{f_{li}}\right)^{n]}_{v_{l}{\bar v}_{i}}\ .}
Here summation over repeated indices is understood;
$\left( Y_{f_{ij}} \right)^{\alpha}_{v_{i}{\bar v}_{j}}$,
$\left(Y_{f_{ij}}\right)^{m}_{v_{i}{\bar v}_{j}}$ are the $f_{ij}$'th
Clebsch-Gordan
coefficients corresponding to the projection of $4\otimes {\Br}_{i}$
and $6\otimes {\Br}_i$ onto
${\Br}_{j}$; and $\Gamma_{\alpha\beta, m}$ is the invariant in
${\bf 4} \otimes {\bf 4} \otimes {\bf 6}$.

Now we will ask whether
the theory is conformal.  As a first step towards proving this let us
show that the
one-loop beta-functions vanish.  For the $i^{th}$ factor $SU(N_i)$
in $G_{{\rm proj}}$, there will generally be Weyl fermions
transforming as:
$$
	\oplus_j a^{\bf 4}_{ij} (N_i, \bar{N}_j)
$$
and scalars transforming as
$$
	\oplus_j a^{\bf 6}_{ij} (N_i, \bar{N}_j)\ .
$$
The one-loop beta-function for the gauge coupling $g_i$
is proportional to
$$
\beta_i  \propto {11\over 3} N_i - {1\over 3} \sum_j (a^{\bf 4}_{ij}
+a^{\bar{\bf 4}}_{ij})N_j
		- {1 \over 2\cdot 3} \sum_j a^{\bf 6}_{ij} N_j\ .
$$
(note that this formula makes sense even if $j=i$ in the sum
because in that case we get an adjoint representation and its second
casimir is $N_i$).
For $N_{i} = Nn_i$ this expression vanishes, because
$$4\cdot N_i=\sum_j a_{ij}^{\bf 4}N_j=\sum_j a_{ij}^{\bar{\bf 4}}N_j$$
$$6\cdot N_i=\sum_j a_{ij}^{\bf 6}N_j$$
and
$${11\over 3}-{2\over 3} 4-{1\over 6} 6=0.$$
Of course this is far from proving that these theories correspond
to conformal
theories. However, considerations of \ksorb\ based on \juanads\
strongly suggest they indeed are.

The considerations above apply to any discrete subgroup
$\Gamma\subset SU(4)$.
However,
if $\Gamma\subset SU(3)$ or $\Gamma \subset SU(2)$
we obtain $\CN=1,2$ respectively.  In these cases it is natural
to consider a reduced quiver which still encodes the theory, taking
into account the corresponding supersymmetry.
For the $\CN=1$ case, we consider the quiver with nodes given
by irreducible representations of $\Gamma$, and consider $k$
arrows from the $i$-th node to $j$-th node if the ${\Br}_j$
representation appears $k$ times in ${\bf 3}\otimes {\Br}_i$.
These correspond to $k$ chiral multiplets $\Phi^{ij}_{f_{ij}}$
in the $(N_i,\bar{N_j})$ representation.
In this case we have a superpotential inherited from the $\CN=4$ theory
which is given by
\eqn\sprpto{\CW = \sum_{i,j,k} \sum_{f_{ij}, f_{jk}, f_{ki}}
h^{f_{ij}, f_{jk}, f_{ki}}_{ijk} {\Tr}\left(  \Phi^{ij}_{f_{ij}}
\Phi^{jk}_{f_{jk}} \Phi^{ki}_{f_{ki}} \right)}
where
\eqn\frml{h^{f_{ij}, f_{jk}, f_{ki}}_{ijk} =
\epsilon_{\alpha\beta\gamma}
\left( Y_{f_{ij}} \right)^{\alpha}_{v_{i}{\bar v}_{j}}
\left( Y_{f_{jk}} \right)^{\beta}_{v_{j}{\bar v}_{k}}
\left( Y_{f_{ki}} \right)^{\gamma}_{v_{k}{\bar v}_{i}}
}
and $Y$'s now correspond to Clebsch-Gordan coefficients for
${\bf 3}\otimes \Br_i\to \Br_j$.  Note that in this context
the vanishing condition for the one loop beta function, $N_f=3N_c$,
follows from the fact that tensoring any representation $N_c$ with
${\bf 3}$
gives a $3N_c$ flavors (and that
if we get adjoints they contribute like $N_c$ fundamentals).

It would be interesting to see whether this vanishing of beta function
persists to higher loops. It is also important to check whether
the anomalous scaling dimensions of the chiral fields are zero.
In fact, the two are related \twoloop\shifvai\robmat.
We now show that the anomalous
scaling dimensions at one loop vanish, which implies together
with the vanishing of the gauge coupling beta functions at one  loop
that they continue to vanish at two loops. The matrix
of anomalous scaling dimensions for chiral fields at one loop
is given by (in our normalizations):
\eqn\anscdm{\gamma_{a\bar d}^{(1)} =  \sum_{b,c} h_{abc} {\bar h}_{dbc} -
2 \delta_{a\bar d}}
where the indices $a,b,c$ are what we called $f_{ij}, f_{jk},  
f_{ki}$, $h_{abc}
\equiv h^{f_{ij}, f_{jk}, f_{ki}}_{ijk}$.

Denote by $W_{ij}$ the multiplicity spaces\foot{We thank
P.~Etingof and D.~Kazhdan
for the explanation of this point}:
${\bf 3}\otimes {\bf r}_{i} = \oplus W_{ij} \otimes {\bf r}_{j}$,
we also have
${\bar{\bf 3}} \otimes \Br_{i} = \oplus W_{ji}^{*} \otimes \Br_{j}$.
The Yukawa coupling $h_{abc}$ may be defined also as follows.
consider the canonical element $\omega = \epsilon^{\alpha\beta\gamma}
e_{\alpha} \otimes e_{\beta} \otimes e_{\gamma}$ of ${\bf 3}^{3}$.
The map $h: \Br_{i} \to {\bf 3}^{3}$, $h(v) = \omega \otimes v$
when expanded in the basis $Y_{a}$ of the spaces $W_{ij}$ etc. has
the following form:
$$
h(v) = \sum_{a,b,c} h_{abc} Y_{a} \otimes Y_{b} \otimes Y_{c} \otimes v
$$
where  $Y_{a} \in W_{ij}, Y_{b} \in W_{jk} , Y_{c} \in W_{ki}$.
Consider the following diagram:
\eqn\dgrm{\matrix{\quad  & & \rho & &  \cr
\quad & \Br_{i} & \longrightarrow & {\bf 3}^{3} \otimes
{\bar{\bf 3}}^{3} \otimes \Br_{i} & \cr
\xi & \downarrow & \swarrow  & \eta & \cr
\quad & {\bf 3} \otimes {\bar{\bf 3}} \otimes \Br_{i} & & & \cr}}
where $\rho (v) = \omega \otimes \bar\omega \otimes v$, for
$v \in \Br_{i}$, $\xi (v) = e_{\alpha} \otimes
{\bar e}_{\alpha} \otimes v$ and $\eta ( a \otimes b \otimes c \otimes
\bar d \otimes \bar e \otimes \bar f \otimes v ) =
\langle \bar d, c \rangle \langle \bar e, b \rangle
a \otimes \bar f \otimes v$.
The identity
$\epsilon^{\alpha\beta\gamma} \epsilon^{\delta\beta\gamma} =
2 \delta^{\alpha\delta}$ implies that
\eqn\idnty{\eta \circ \rho = 2 \xi.}
Now let us rewrite \idnty\ in the basis $Y_{f_{ij}}$ of intertwiners.
We shall use the indices $a$ for $f_{ij}$, $b$ for $f_{jk}$ and $c$ for
$f_{ki}$.
First of all, it is easy to represent the map $\rho$ as:
$$
\rho (v) = \sum_{a,b,c,d,e,f} h_{abc} {\bar h}_{def}
Y_{a} \otimes Y_{b} \otimes Y_{c} \otimes
{\bar Y}_{f} \otimes {\bar Y}_{e} \otimes {\bar Y}_{d} \otimes v
$$
where $Y_{a} \in W_{ij}, Y_{b} \in W_{jk} , Y_{c} \in W_{ki},
{\bar Y}_{d} \in W_{mi}^{*}, {\bar Y}_{e} \in W_{nm}^{*}, {\bar  
Y}_{f}^{*}
\in W_{in}^{*}$ and we sum over $j,k,m,n$.
Under the $\eta$ map the
only vectors which are not mapped to zero are those
for which $b=e$, $c=f$, hence $m=k$, $n=j$, hence
$$
\eta \circ \rho (v) = h_{abc} {\bar h}_{dbc} Y_{a} \otimes {\bar Y}_{d}
\otimes v
$$
On the other hand it is equal to $2 Y_{a} \otimes {\bar Y}_{a}\otimes v$,
hence $\gamma_{ad}^{(1)} = 0$.

If $\Gamma \in SU(2)$ we get an $\CN=2$ theory.  In this case
the nodes of the quivers correspond to irreducible representations
of $\Gamma$ and links correspond to the decomposition of representation
upon tensoring with ${\bf 2}$.  This gives rise to the well known
affine A-D-E Dynkin diagrams where the $n_i$ correspond to Dynkin
indices associated with each node.  In fact these
$\CN=2$ theories have been studied in \kmvgeom\ where it was shown
that they are the {\it only} conformal $\CN=2$ theories
if the gauge group is a product of $SU$'s and the matter is in the
bifundamental
representations.  Note that the $2$ in the conformality condition
$N_f=2 N_c$
is
the dimension of the fundamental
representation of $SU(2)$.  In fact, the power of orbifolding in this
case suggests that perhaps even in the $\CN=1$ and $\CN=0$ cases
considered above, in the subclass of gauge theories corresponding
to product of $SU$ groups with matter in bifundamentals we have a full
class of allowed conformal theories.  It is quite interesting
that discrete subgroups of $SU(3)$  \suthree\ and $SO(6)$ \sosix\
have already been
classified. For example, for the $SU(3)$ case, in addition to the
subgroups of $SU(2) \times U(1)$ one has: two infinite seria
$\Delta (3n^2)$
and $\Delta (6n^2)$ (which are analogues of $A_n$ and $D_n$'s in $SU(2)$
case - they are extensions of $\IZ_{n} \times \IZ_{n}$);
and six exceptional cases $\Sigma (d)$,
$d=60,168,360\varphi,36\varphi,
72\varphi,216\varphi$, where $\varphi =1$ or $3$ depending on whether
the group belongs to $SU(3)/{\IZ}_{3}$ or $SU(3)$ respectively. The
number in braces is the order of the group.

It should be noted that a generic choice of $N_i$s
would not lead to a conformal theory.
A conceptually useful example of a projection which does
not lead to a superconformal theory is one for which the
Chan-Paton factors transform in $n$ copies of a single irrep
${\bf r_1}$ of $\Gamma\subset SU(2)$.  It is easy to see that there
will be
no hypermultiplets after the projection
and the theory will be a pure $SU(n)$ $\CN=2$
gauge theory, which is not superconformal.

\newsec{Relations to string theory}
If we realize these field theories via orbifolds of D3-brane theories,
different
choices of representations for the Chan-Paton factors will
have definite physical meanings \dmquiver\polten\dougegs.
If we wish to describe $N$ D-branes away from the fixed point,
the unprojected theory will be $U(N \vert \Gamma \vert)$
and the indices $I,J$ will lie in $N$ copies of the regular
representation of $\Gamma$.  This is in fact why we considered
the action of $\Gamma$ in the previous section.
Other representations will involve Chan-Paton factors living at the
fixed point of the orbifold; they will correspond to D5- or D7- branes
wrapped around shrunken 2- or 4-cycles \polten\dougegs .  Given
the fact that these other physical projections make sense, it remains
to explain why not all the orbifoldings give rise to conformal theories
from the supergravity viewpoint \juanads .

\newsec{Moduli Space of Couplings in the Conformal Theory}
So far we have only considered the action induced from the $\CN=4$
theory.  This means that the conformal theory has at least one free
coupling constant, which is that inherited from $\CN=4$.
It is natural to ask whether there are other deformations of
this theory.  In the case of $\CN=2$, which corresponds to the ADE
examples noted before, it is known that there is a moduli space of
deformations, one coupling for each group.  It was suggested in \ksorb\
that this may be related to the blowup modes also expected from the
supergravity orbifolds.  As noted in
\ksorb\ there is a puzzle here as one would naively
expect the blow-up modes to correspond to FI-terms.
However we believe that in the present context 3-branes are  
secretly 5-branes
wrapped around the
extra vanishing two spheres, as has been suggested
in \polten\dougegs\frbranes, in which
case the couplings will be related to the expectation
values of NS-NS $B_{ij}$ (related to coupling constant of gauge theory)
and RR $B_{ij}$ (changing the theta angle) \dmquiver\ fields.

As further confirmation of this picture, we now connect this  
picture to the
known result that the
moduli space of couplings for these theories is the moduli
space of flat $ADE$ connections on the torus \kmvgeom (for the $A$
case see also \mfour).  Let us first discuss more explicitly
this moduli space.
Let us consider flat $G$-connection $A_{z, \zb}$ on a two-torus
$E_{\tau} = \IC / \left( {\IZ} + {\tau}\IZ \right)$ with complex
parameter $\tau$. Let $r$ be the rank of the group $G$ and let   
$H_{i}$'s $i=1,
\ldots, r$
be the generators of the Cartan
subalgebra of $G$. It can be shown that by a complex gauge  
transformation the
$(0,1)$
component $A_{\zb}$ can be brought to a constant  
$\liet^{\IC}$-valued form:
$A_{\zb} = \sum_{i} {{\tau_{i}}\over{\tau - {\bar \tau}}} H_{i}$.
Note that the $\tau_i$ are defined up to the
action of the Weyl group of $G$.  Moreover the gauge  
transformations of the
form
\eqn\gge{
g ( z, \zb) =
\exp \left({{2\pi i}\over{\tau - \bar \tau}} \left( ( z- \zb) a_{i} + ( z
\bar\tau - \zb \tau ) b_{i} \right) H_{i} \right),  \quad a_{i},  
b_{i} \in \IZ}
 shift the value of $\tau_{i} $ by $a_{i} + b_{i} \tau$.
We can introduce an extra $\tau_{0}$, obeying the property:
$\sum_{i=0}^{r} n_{i} {\tau}_{i} \equiv 0 {\rm mod} \left( {\IZ} +  
{\tau} \IZ
\right)$ (where $n_0=1$).
This presentation makes the action of $SL_{2}({\IZ})$ of the underlying
torus manifest:
\eqn\sltwo{
\tau \to -1/{\tau}, \quad  {\tau}_{i} \to {\tau}_{i}/{\tau}.}

Now we connect this to the moduli that we expect to see in the
orbifold picture with the branes.
By writing
$$
\tau_{i} = x_{i} + y_{i} \tau,  \quad x_{i}, y_{i} \in \IR, i = 0,  
\ldots , r
$$
with the condition
$\sum_{i} x_{i} \equiv \sum_{i} y_{i} \equiv 0\  {\rm mod}\ {\IZ}$  
we make the
connection to the fluxes of the $B$-fields  explicit:  
$\int_{\Sigma_{i}} B^{NS}
= y_{i}$,
$\int_{\Sigma_{i}} B^{RR} = x_{i}$.  Here $\Sigma_{i}$ are the two-cycles
of the corresponding ALE space, which are in one-to -one  
correspondence with
simple roots of $G$.
The action of $IIB$ $S$-duality group on $B^{NS}, B^{RR}$
and string coupling $\tau$ translates to the action \sltwo. We may  
also think
of the $\tau_{i}$'s
as the points on the torus $E$ (actually, on the dual torus, which  
coincides
with $E$ as complex
manifold). The periodicity
of the torus $x, y \sim x, y + 1$
becomes the periodicity of the $B$-fields.
The $\tau_{i}$ is  the coupling for the $SU(Nn_i)$ factor.
The orbifold point corresponds to $x_{i} = 0$, $y_{i} =  
{{n_{i}\over{\vert
\Gamma \vert }}}$.  Note that the Weyl group action
in the gauge theory setup is exactly mapped to the monodromy
action in the context of ADE singularities.  We thus see they
are indeed identical.

It would be interesting
to extend this to the case of $\CN=1$ and $\CN=0$,
where again we may expect the 3-brane to be secretly
a combination of 5-branes and 7-branes.  The moduli of the corresponding
orbifold theory is expected to be related to the analogs of the various
$B$ fields
and the 4-form gauge field expectation values.  This should be  
interesting
to develop and reconcile with the supergravity picture.

Another issue
is the global identifications of the moduli space.
This
will correspond to non-trivial dualities in the field theory setup,
which in the $\CN=2$ is completely known.
Since we do not know the full description of the moduli for the
 $\CN=1$ and $\CN=0$ cases we cannot specify the full duality group.
However, there is already one complex modulus of the conformal theories
which is inherited from the $\CN=4$ coupling $\tau$
(which as noted before is related to the coupling of the gauge groups by
$\tau_i=\tau {{n_i}\over{\vert \Gamma \vert}}$).  In this case the  
$SL(2,\IZ)$
invariance
of Montonen-Olive will give rise to $SL(2,\IZ)$ symmetry in this  
subspace.
In other words we thus expect to have a {\it self-duality} in this
direction of coupling constant space.  This is quite exciting, given
that we are dealing with field theory S-dualities including the  
$\CN=0$ case.

If our conjecture about the vanishing of beta functions
in the $\CN=0$ is true, this would indeed inovlve an amazing
set of cancellations which is not a direct result of supersymmetry.
It is natural to expect applications of this idea
in the context of hierarchy problem.
For example, in all of the models we have it is easy to see that
we have equal numbers of bosons and fermions, even though we do not
have supersymmetry.
This is quite intriguing and may point to a new
physical symmetry principle.

\bigskip
\centerline{{\bf Acknowledgements}}

We would like to thank O. Bergman, P.~Etingof,
A. Johansen, D.~Kazhdan, E.~Silverstein and A.
Strominger for
valuable
discussions.  We are also grateful to Z. Kakushadze for pointing out the
reference \iba\ as well as for discussions on the beta functions
and to B.~Zwiebach for bringing \hps\ to our attention.

This work was supported in part by
NSF grant PHY-92-18167.  In addition
the research of N. N. was supported by Harvard Society of Fellows,
partially by RFFI under grant 96-02-18046 and partially
by grant 96-15-96455 for scientific schools.

\listrefs
\bye